# High Density Out-of-Plane Microprobe Array

[1]C.H. Huang, [1]C. Tsou, and [2]T.H. Lai

[1]Department of Automatic Control Engineering, Feng Chia University,
100 Wenhwa Road, Seatwen, Taichung, Taiwan, 40724 R.O.C.

[2]The Graduate Institute of Electrical and Communications Engineering, Ph.D. Program,
Feng Chia University, 100 Wenhwa Road, Seatwen, Taichung, Taiwan, 40724 R.O.C.

*Abstract*-In this paper, the high density out-of-plane microprobe array is demonstrated. The fabrication processes of proposed device including bulk micromachining, thin film deposition and electroplating. By depositing the various thickness of thin film with residual tensile stress, the deflection of bending beam could be precisely controlled. The horizontal variation of bending beam array could also restrict in a few micrometer. The concept for the device fabrication is depositing the Ti thin film on a suspension $SiO_2$ cantilever to bring out-of-plane bending deformation, and then use electroplating process such as low stress Ni film to increase the stiffness of the probe structure without change the beam's bending profile. Through the different electroplating bath, the bending microprobe with different material, such as Ni and Cu, could be fabricated for IC testing and interconnection applications. According to the mask design, the high density probe array could easily achieve. The prototypes of the 40×40 microprobe array in pitch sizes of 50μm, 80μm and 100μm were fabricated and characterized. Measurement result by the optical interferometer shown the mean bending height of twenty samples are about 3μm, 9μm and 26μm in different beam length 25μm, 55μm and 85μm, besides the maximum horizontal variations are less than ±0.8μm, ±1.2μm, ±2μm that corresponded to the pitch size of 50μm, 80μm, and 100μm, respectively.

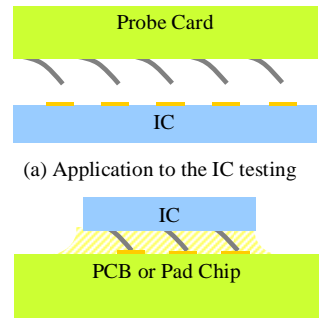

(a) Application to the IC testing

(b) Application to the flip chip interconnect

Fig.1 The applications of the out-of-plane bending cantilever beam.

## I. INTRODUCTION

MEMS technology has been developed rapidly in the last few years. More and more special micro structures were discussed in several publications. However, all of the structures were produced by consisting of the three fundamental structures, which included bridge, cantilever and membrane structures. Even the more complex structures were no exception. The cantilever with the property of simple design and easy fabrication among three kinds of fundamental structure; therefore, it was popular used in the design of MEMS device. In this paper, the cantilever beam used as the probe microstructure has been achieved with high density and out-of-plane bending deformation. The proposed microprobe array could widely applied in IC testing [1-4] and flip chip interconnect [5-9], as shown in Fig. 1(a) and Fig. 1(b), respectively.

Due to the develop roadmap of the IC chip, the scale of line width were decrease rapid in every generation. Therefore, the technology of the fabrication, package and test were changed in order to compatibility with the IC development. In the test technology, the probe card plays an important role in the IC test. The property of fine pitch and the high pin count probe array will the trend of the IC development. Hence the traditional probe card which assembled by the manual were facing the bottleneck. Utilizing the out-of-plane bending structure fabricated by the MEMS technology could easily achieve these restrictions. Because of the out-of-plane microprobes could easy arrange in the fine pitch and the demand shape by the mask design. Consequently, the out-of-plane microprobe had the great potential to replace the traditional probe card. Besides, these out-of-plane microprobes also could use as the flip chip interconnects. As a result of the scale down to the micrometer degree, the scale factor will be concern in many physical phenomena. Particularly, the thermal expansion was one kind of the physical phenomena which were produce thermal mismatch between the flip chip and package substrate at the micrometer scale degree. Hence, the traditional flip chip interconnect of the solder ball will face to the challenge. Utilizing the out-of-plane microprobe array as the flip chip interconnect could overcome this problem. The thermal stress was released through the deformation of the bending structure. Therefore, it will be the candidate for the next generation interconnects of the flip chip.

Utilizing the out-of-plane bending structure as the microprobe or interconnect, the electrical or mechanical property will be concern in the every report but no one discusses the horizontal variation in each probe. However, the horizontal variation in each probe will be the important property along with the increasing probe number. If the probe array with the minima horizontal variation in each probe, the requirement of the mechanical property will be not so strict. Therefore, this paper focus on the utilizing of the simply and cheap process to manufacture the out-of-plane microprobe array which have the minimal horizontal variation.

There are several MEMS approaches reported on the fabrication of out-of-plane bending structure. Most of all fabrications of the out-of-plane structure were by the two layers material with different stress [2-4], but those methods usually have some disadvantage. First, the structure wasn't robust enough and couldn't load with larger force than single layer. Supposing the bi-layer structures load a force, it was easily peeling between two layer interfaces. Second, these bi-layer structures have much material restrict in the fabrication. In order to obtain bending up structure, the material must be concern seriously. Therefore, the fabricate process were restrict from the material choice. In this paper, the simple fabrication was provided to achieve high density out-of-plane probe array with single layer. By depositing a tensile stress metal film on a suspension beam, the out-of-plane bending structure was formed. As such, the bending deformation could be controlled by the thickness of tensile stress metal. Moreover, a minimal horizontal variation in each microprobe could obtain. Furthermore, the electroplating process was employed to form the main probe structure. Utilizing this fabrication process not only obtain better mechanical characteristic but also have more flexibility in material choose, such as Cu and Ni-Co probe array.

## II. Design

The height density out-of-plane microprobe array was designed and demonstrated. The design concept of the out-of-plane microprobe array was shown in Fig.2. The fabrication technology of proposed device included bulk micromachining, thin film deposition and electroplating. The characteristic of the out-of-plane microprobe array can be controlled by the deposition thickness and materials of the thin film. For example, the metal film such as aluminum (Al) by evaporation has the compression residual stress but titanium (Ti) brings tensile residual stress and the stress magnitude varies with the thickness of metal film. However, if the material with large tensile stress, such as chromium (Cr), used as a stressed cover layer, the cantilever beam will roll up or break. Therefore, the choused material was very important in the thin film deposition for the bending microprobe. In this paper, the Ti is used as the stressed material which cause the suspension microprobe bend up. By controlling the deposition thickness, the out-of-plane deformation of each microprobe could restrict in a few micrometer. After the pre-bending structure was fabricated by deposit thin film process, the electroplating process could used to increase the stiffness of the probe microstructure. In this paper, the electroplated nickel film was used as the pbobing structure. The nickel fabricated by electroplating process usually has lowered residual stresses than others materials. Therefore, it affects the deformation of the pre-bending probe slightly. Besides, the single layer structure could avoid some problems which will occur on the bi-layer structure such as peeling between two layer interfaces. Hence, the life-time of the microprobe could be increase.

In the mask design, the probe tip could design in different shape to satisfy different applications. For instance, the probe tip design in triangle shape could use in IC testing that pads made from Al or Cu has native oxide layer. The tip with triangle shape will effectively break the oxide film. In the other hand, when the probe tip design in the flat tip, the probe could use as the interconnect in the IC packaging. The tip with flat shape will increase the contact area with the substrate. Therefore, base on those different applications, the probe will design according to the user's requirement. The prototype of the micro probe array is designed in the three specifications, the pitch between the microprobe arrays are about 50μm, 80μm and 100μm respectively. The detail geometry parameter correspond to different pitch size between etch other is shown in table1.

In this paper, we emphasize utilizing the simple and reliable process to manufacture the microprobe array with minimal horizontal variation. Therefore, the prototype is not concern in detail such as electrical isolation between each probe and external environment electrical connecting; we just focus on the microprobe structure itself. However, we already have some ideal to overcome these problems. Utilizing the through-hole electrode, the microprobe could connect to external environment. In the order hand, the laser cutting process could isolate the electrical between each microprobe. According to the above mention methods, it could completely realize the high density, fine pitch commercial microprobe array.

## III. Fabrication

The out-of-plane structure array is fabricated with bulk machine and thin film deposition techniques. First step was the fabrication of suspended cantilever beam by photolithography and bulk etching process. Then deposit a metal with tensile stress on the cantilever beam to form the pre-bending structure. Finally, the electroplating process is used to enhance the stiffness of the structure. The detail process flow as show in Fig.3 and describe in the following statements. First, the thermal oxide with 1μm thickness was grown on the silicon substrate (Fig. 3a). Next, define the release region around out-of-plane structure by photolithography process and oxide was wet etched by buffer oxide etcher (BOE) (Fig. 3b). Utilize oxide layer as the hard mask to etch the silicon substrate. The silicon substrate was etched by consistency 25% KOH solution at 75°C (Fig. 3c). Then evaporate the Ti with 4000Å thickness on the suspension cantilever beam (Fig. 3d). The cantilever beam was bending upward due to the tensile residual stress in titanium layer. Next, the nickel was electroplated to enhance the stiffness of the structure (Fig. 3e). Finally, remove the oxide layer and etch titanium metal layer under the nickel cantilever beam to obtain the single layer structure (Fig. 3f).

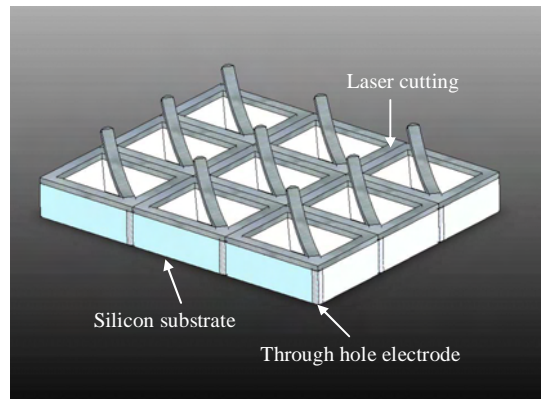

Fig.2 The design concept of out-of-plane microprobe array
Table1.
The different pitch of microprobe array  correspond to geometry parameter

| pitch(μm) | Beam-length(μm) | Beam-width(μm) |
|---|---|---|
| 50 | 25 | 6 |
| 80 | 55 | 10 |
| 100 | 86 | 10 |

In the step (d), if we do not execute next process rapidly, the Ti expose in the air environment will grow a native oxide layer and affect the follow up electroplating process. Therefore, the nickel (Ni) could use to deposit on the Ti layer as the seedlayer for following electroplating process. The nickel layer depositing by evaporation have low residual stress characteristic. Consequently, when the nickel deposit on the pre-bending beam, it'll not be affected the pre-banding height so much.

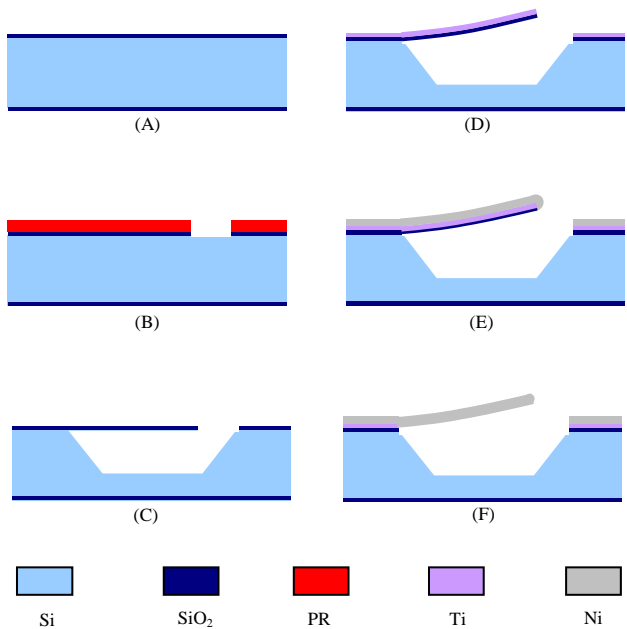

Fig.3 fabrication flow : (a) Thermal oxidation on the silicon substrate (b) Pattern the probe area by photolithography (c) Silicon substrate is etched by KOH (d) The Ti is evaporated on the suspending oxide cantilever (e) Electroplating Ni (f)the oxide and Ti layer is removed by BOE

## IV. RESULT AND DISCUSSION

The scanning electron microscope (SEM) image of the fabricated microprobe structure with pitch size 50μm, 80μm and 100μm are shown in Fig. 4(a), 4(b) and 4(c), respectively. According to the SEM image, the bending height with excellent horizontal uniformity was observed. The deflection of each out-of-plane microprobe is determined by the optical interferometer and the measurement result shown in Fig. 5(a), 5(b) and 5(c). The result shows the bending height are 3μm, 9μm and 26μm corresponding to the pitch 50μm, 80μm and 100μm, respectively.

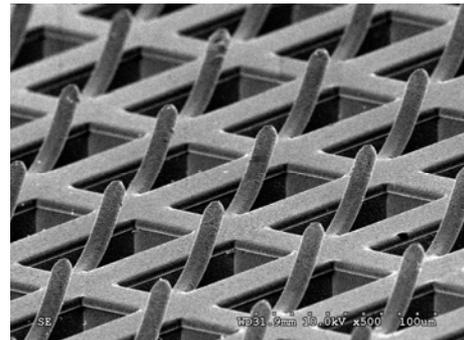

(a) The SEM image of the microprobe array under the pitch 100μm

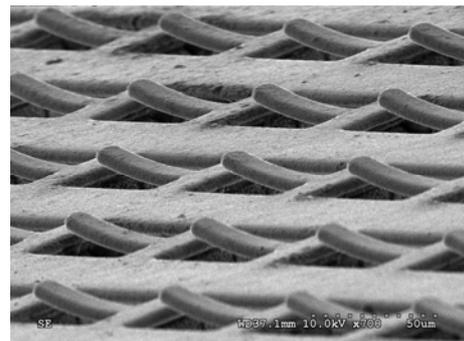

(b) The SEM image of the microprobe array under the pitch 80μm

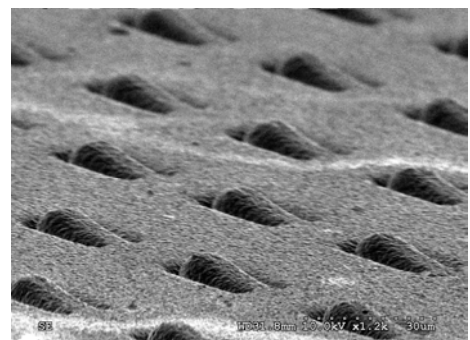

(c) The SEM image of the microprobe array under the pitch 50μm

Fig.4 The SEM image of the microprobe array under the different pitch size

In order to investigate the uniformity in each array which consist of 1600 (40×40) microprobes. The average value is obtained through the sampling twenty points in each specification. Figure 6 shows the measurement result about the relationship between the beam length and bending height. The error bars in this figure represent the deflection variation. According to the Fig. 6, we could find some phenomenon in the experiment result. The beam length is in direct proportion to the bending height and the horizontal variation is also increasing along with the beam length. Because of the stiffness of the microprobe affect the horizontal variation in each array with different specification. The microprobe has the longer beam length, the lower stiffness is achieved. Therefore, the more horizontal variation is caused by the residual stress of the thin film. The prototype in each specification show the deflection variation are ±0.8μm, ±1.2μm and ±2μm corresponding to the pitch of microprobe array 50μm, 80μm and 100μm, respectively.

The microprobe array with different material could easy fabrication by the different electroplating bath in the same process that provided in this paper. Fig. 7 shows the microprobe array fabricated by the Cu. Although the Cu is used to fabricate the microprobe, the characteristic of the excellent uniformity also exist. When the microprobe used in the IC test, the material of Ni is chosen as the main structure. Because of the nickel have more stiffness and also could endure the higher contact force than others. In the other hand, if the material of copper is chosen as main structure it is suitably use in the interconnect due to the lower resistance. Therefore, the microprobe array can made from different materials according to the various applications.

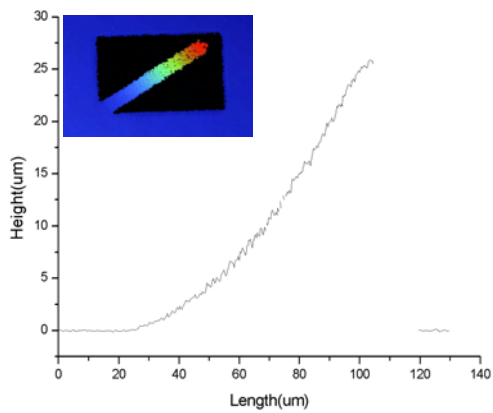

(a) The measured result under the pitch 100μm

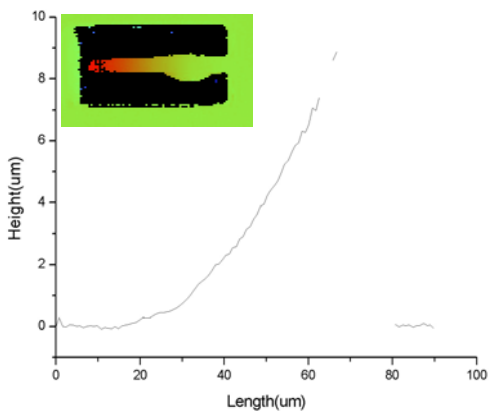

(b) The measured result under the pitch 80μm

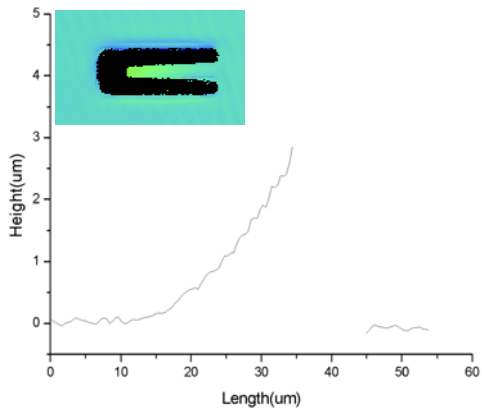

(c) The measured result under the pitch 50μm

Fig5.The bending height value of the different microprobe array measured by optical interferometer

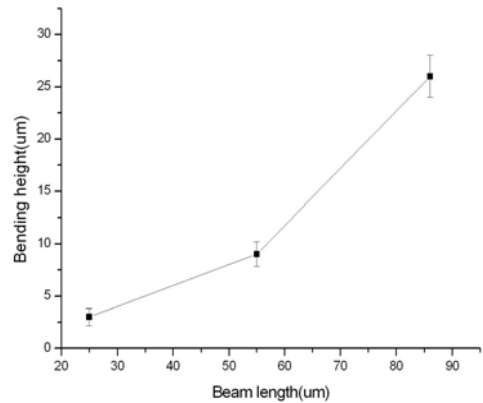

Figure 6. The relationship between the beam length and bending height

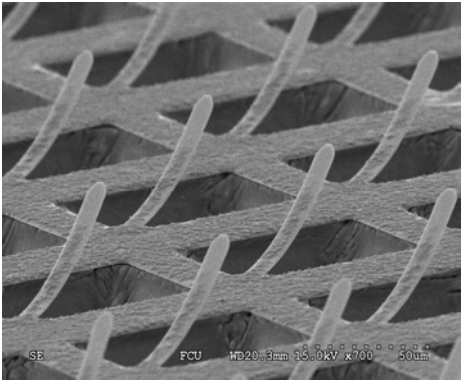

Figure 7. SEM image of the out-of-plane microprobe array with copper electroplated

## V. CONCLUSION

In this paper, the high density out-of-plane microprobe arrays with excellent uniformity are demonstrated. The variation of out-of-plane deflection is ±0.8μm, ±1.2μm, and ±2μm corresponding to the pitch 50μm, 80μm, and 100μm, respectively. According to the fabrication process of the microprobe, we can use the effect of residual stresses of thin films deposition to bent the flexible micromachined probe with a large out-of-plane predeformation and combine post-electroplating technique to further enhance its stiffness. Besides, it is clear that using the process reported here can fabricate a microprobe array with the characteristics of both lower residual stresses and minimum deflection variation under large out-of-plane deformation. This technique has the potential for achieving high production yield with fine pitch, low cost, high pin counts and area array testing as well as flip chip packaging.

ACKNOWLEDGMENT

The authors would appreciate the financial support by Taiwan Semiconductor Manufacturing Company (TSMC), and the fabrication process support by Precision Instrument Support Center of Feng Chia University and the National Nano Device Laboratory (NDL) in providing the fabrication facilities.